\documentclass[prc,aps,superscriptaddress,floatfix,twocolumn,nofootinbib]{revtex4-2}
\usepackage{graphicx}
\usepackage{dcolumn}
\usepackage{bm}
\usepackage{CJK}
\usepackage{amssymb,amsmath}
\usepackage{makecell}
\usepackage{multirow}
\usepackage{appendix}
\usepackage[colorlinks,
            linkcolor=red,
            anchorcolor=blue,
            citecolor=green
            ]{hyperref}
\begin{document}

\title{Sub-barrier d+$^{208}$Pb scattering and sensitivity to nucleon-nucleon interactions}
\author{Peng Yin}
\thanks{Corresponding author: pengyin@iastate.edu}
\affiliation{College of Physics and Engineering, Henan University of Science and Technology, Luoyang 471023, China}
\affiliation{CAS Key Laboratory of High Precision Nuclear Spectroscopy, Institute of Modern Physics, Chinese Academy of
Sciences, Lanzhou 730000, China}
\affiliation{Department of Physics and Astronomy, Iowa State University, Ames, IA 50011, USA}
\author{Weijie Du}
\thanks{Corresponding author: duweigy@gmail.com}
\affiliation{Department of Physics and Astronomy, Iowa State University, Ames, IA 50011, USA}
\author{Wei Zuo}
\affiliation{CAS Key Laboratory of High Precision Nuclear Spectroscopy, Institute of Modern Physics, Chinese Academy of
Sciences, Lanzhou 730000, China}
\affiliation{School of Nuclear Science and Technology, University of Chinese Academy of Sciences, Beijing 100049, China}
\author{Xingbo Zhao}
\affiliation{CAS Key Laboratory of High Precision Nuclear Spectroscopy, Institute of Modern Physics, Chinese Academy of
Sciences, Lanzhou 730000, China}
\affiliation{School of Nuclear Science and Technology, University of Chinese Academy of Sciences, Beijing 100049, China}
\affiliation{Advanced Energy Science and Technology Guangdong Laboratory, Huizhou, Guangdong 516000, China}
\author{James P. Vary}
\affiliation{Department of Physics and Astronomy, Iowa State University, Ames, IA 50011, USA}

\begin{abstract}
We employ the non-perturbative time-dependent basis function (tBF) approach to solve for the sub-Coulomb barrier scattering of the deuteron projectile on the $^{208}$Pb target. Specifically, we treat the target as a source of a strong external Coulomb field that induces higher-order effects in electric-dipole excitations of the deuteron projectile including the population of states not accessible through direct electric-dipole transitions from the ground state of the deuteron. We calculate the electric-dipole polarizability of the deuteron and elastic scattering observables for comparison with experimental data. With no adjustable parameters, the tBF approach provides good agreement with experimentally available differential cross-section ratios. The dependence of these measured quantities on nucleon-nucleon interactions is investigated. We also investigate the detailed scattering dynamics and identify characteristics of coherent and incoherent processes.
\end{abstract}

 \maketitle

\section{Introduction}
As the existing and upcoming radioactive ion beam facilities provide unique access to the exotic phenomena in nuclear physics, predictive and reliable reaction theory is crucial for establishing a substantial connection between the measured reaction observables and the underlying nuclear structure. While phenomenological reaction theories and standard approximation methods are playing a significant role in modeling nuclear reaction data, reaction approaches, that start from the {\it ab initio} calculations of nuclear structure, have been developed and turn out to be numerically tractable and successful in describing reactions involving light nuclei. For few-body systems with nucleon number $A\le 4$, the Faddeev~\cite{Witala:2000am}, Faddeev-Yakubovsky~\cite{Lazauskas:2004hq,Lazauskas:2009gv}, hyperspherical harmonics (HH)~\cite{Marcucci:2009xf,Marcucci:2019hml}, the Alt, Grassberger, and Sandhas (AGS)~\cite{Deltuva:2006sz,Deltuva:2007xv}, resonating group method (RGM)~\cite{Hofmann:2005iy}, etc.,  are applicable and successful. For systems with more than four nucleons, additional approaches, such as the Lorentz integral transform (LIT) methods~\cite{Gazit:2005nh,Quaglioni:2007eg,Bacca:2008tb,Leidemann:2012hr}, the Green's function Monte Carlo (GFMC)~\cite{Nollett:2006su}, the fermionic molecular dynamics (FMD)~\cite{Chernykh:2007zz}, no-core shell model (NCSM) with RGM~\cite{Quaglioni:2008sm,Quaglioni:2009mn,Navratil:2009ut,Navratil:2011zs} and the single-state harmonic oscillator representation of scattering equations (SS-HORSE) method~\cite{Shirokov:2018nlj} have been proposed and applied. However, these approaches may be challenged to retain the full, non-perturbative quantal coherence of all the potentially relevant intermediate and final states in nucleus-nucleus reactions, especially for reactions involving unstable rare isotopes.

In Refs.~\cite{Du:2018a,Du:2018b}, we proposed the theoretical framework of the non-perturbative time-dependent basis function (tBF) approach. The tBF method establishes a firm connection between {\it ab initio} nuclear structure approaches, e.g., the NCSM~\cite{Barrett:2013nh}, and nuclear reaction theory. Its counterpart in relativistic quantum field theory, the time-dependent basis light front quantization (tBLFQ) approach, has been applied to investigate the non-linear Compton scattering~\cite{Zhao:2013cma,Hu:2019hjx}, the interaction of an electron with intense electromagnetic field~\cite{Chen:2017uuq} and the scattering of a quark on a heavy nucleus~\cite{Li:2020uhl,Li:2021zaw} at high energies. The key point of the tBF method is the construction of the basis representation for the scattering problem, employing the eigenstates of the system being investigated (a rare isotope beam in future applications) that are obtained with an {\it ab initio} nuclear structure approach. Within the basis representation, we solve the equation of motion (EOM) of the scattering process numerically as an initial value problem in a non-perturbative manner, where the quantal coherence is fully retained during the reaction process. It is noteworthy that the tBF approach appears to be well-suited as a forefront application of quantum computing in nuclear scattering dynamics~\cite{Du:2020glq} just as its relativistic counterpart, tBLFQ, has been shown recently to be well-suited for quark jet scattering applications on quantum computers~\cite{Wu:2024adk}.

In Ref.~\cite{Yin:2019kqv}, we improved the tBF method and employed it to investigate the deuteron scattering on $^{208}$Pb well below the Coulomb barrier of approximately $11$ MeV. By considering all the possible electric-dipole ({\it E1}) transition paths among all the states in the deuteron projectile and the polarization potential, we successfully described the experimental data of elastic cross-section ratios with the tBF method employing a realistic nucleon-nucleon ({\it NN}) interaction from the chiral effective field theory (EFT). No adjustable parameters were introduced. In the present work, we study the sensitivity of the deuteron {\it E1} polarizability and the d+$^{208}$Pb scattering to the choice of the {\it NN} interaction. We also investigate the detailed scattering dynamics.

The paper is organized as follows. In the next section, we present the theoretical framework of the tBF approach. We show the main results in Sec.~\ref{sec:resultsAndDiscussion}. Finally we give a summary of our conclusions and an outlook in Sec.~\ref{sec:conclusions}.

\section{Theoretical framework of the tBF approach}
\label{sec:theoryReview}

In this work we investigate the scattering of the deuteron projectile on the $^{208}$Pb target below the Coulomb barrier employing the tBF approach. When the Sommerfeld parameter \begin{equation}
\eta=\frac{Ze^2}{v}
\label{eq:sommerfeld}
\end{equation}
is much larger than unity, one may reasonably adopt a semi-classical approach for the scattering trajectory~\cite{Alder:1956im,Alder:1975}. Here $Z$ represents the charge of our target ($Z = 82$ for Pb), $e$ is the unit of electric charge and $v$ represents the incident velocity of the deuteron. Note that we adopt natural units and set $\hbar=c=1$ throughout the paper. The minimum value of the Sommerfeld parameter in this work is $6.93$ (corresponding to the maximum incident deuteron energy $7$ MeV). Therefore we assume, for the applications in the present work, that the center of mass (COM) of the deuteron projectile moves along a classical trajectory. We then construct the time-dependent Schr\"{o}dinger equation of the scattering process describing how the external field provided by the $^{208}$Pb target induces the internal transitions in the deuteron projectile during the scattering. The main idea of the tBF approach is to solve the time-dependent Schr\"{o}dinger equation within a basis representation determined from {\it ab initio} nuclear structure calculations.

We do not consider the excitation of $^{208}$Pb since its spin-parity ($3^-$) requires a high multipole ({\it E3}) transition, its excitation energy ($2.6$ MeV) is also rather high~\cite{Aoki:2000nim} and the deuteron provides an external excitation potential acting on the lead target that is $82$ times weaker than it experiences from the lead target. Hence, this $3^-$ excitation of $^{208}$Pb is expected to be at least two orders of magnitude smaller than the Coulomb dissociation of the deuteron for $E_d=3-7$ MeV, as previously noted~\cite{Rodning:1982zz}. The influence of vacuum polarization, atomic screening and relativistic corrections on the elastic cross-section ratios is found to be small~\cite{Moro:1999fcl,Rodning:1982zz} and therefore these effects are also not taken into account in this work. We found the effects of the magnetic dipole ({\it M1}) transitions to be negligibly small compared to the effects of the {\it E1} transitions below the Coulomb barrier and would not affect the conclusions of this paper so we omit the {\it M1} transitions at the present time~\cite{Yin:2019kqv}.

\subsection{Determination of the trajectory}
\begin{figure}[tbh]
\begin{center}
\includegraphics[width=1.0\columnwidth]{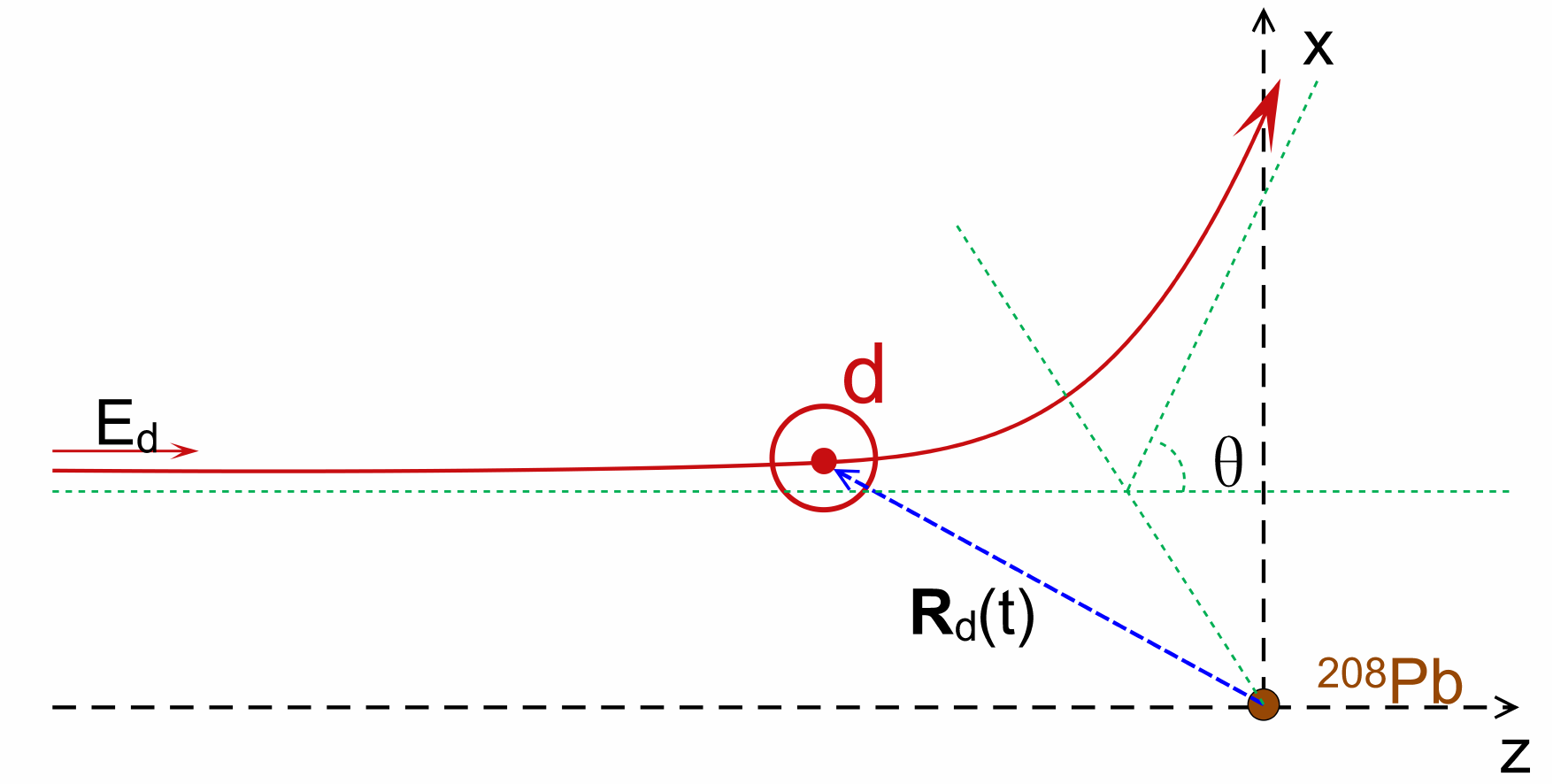}
\end{center}
\caption{(Color online) A sketch for the scattering of the deuteron projectile on the $^{208}$Pb target. See the text for details.}\label{fig1}
\end{figure}

We show in Fig.~\ref{fig1} the sketch of the scattering setup. We set the scattering plane to be the $xz$ plane. We assume that the $^{208}$Pb target is infinitely massive and we then work in the lab frame. For the low incident energies considered, we assume that the $^{208}$Pb target is a point-like nucleus located at the origin during the scattering. The initial velocity of the deuteron projectile is parallel to the $z$ axis. $E_d$ and $b$ denote the bombarding energy of the deuteron and the impact parameter, respectively. $\theta$ represents the scattering angle. The time-dependent vector ${\bm R}_d(t)$ denotes the position of the COM of the neutron-proton ({\it np}) system with respect to the origin during the scattering. We calculate the trajectory assuming that the deuteron projectile is initially located at $(x,z)=(b,z_0)$ at $t=0$ as shown in Fig.~\ref{fig1}. Specifically, we adopt $z_0=-300$ MeV$^{-1}$ ($\approx -59200$ fm) for which the scattering observables investigated in this work are converged to the order of $10^{-5}$~\cite{Yin:2019kqv}.

 In the non-relativistic limit, the trajectory for the scattering of two electric point charges is given by the well known Rutherford trajectory. When two nuclei approach to distances comparable to the sum of their radii, they interact through the strong nuclear potential and the trajectory will deviate from the Rutherford trajectory. Even for scatterings well below the Coulomb barrier, where the nuclear effects become small, some other effects will also lead to deviations from the Rutherford trajectory. One of the most important effects is expected to arise from dipole polarization~\cite{Moro:1999fcl}. In this work, the effect of the dipole polarization will be represented by a polarization potential acting on the COM of the deuteron during the scattering process. Simultaneously, we will treat the effects of the external field on the relative motion inside the deuteron at the quantum amplitude level in tBF.

Taking into account the dipole polarization effect of the deuteron projectile, the trajectory of the projectile is determined by the following combined potential
\begin{eqnarray}
V_{\rm pot} &=& V_{\rm c}+V_{\rm pol},
\label{eq:Vpot}
\end{eqnarray}
where $V_{\rm c}$ and $V_{\rm pol}$ denote the Coulomb potential and the polarization potential, respectively, acting on the projectile COM. When omitting the correction of the polarization potential, the trajectory takes the form of the conventional hyperbolic Rutherford trajectory.

Applying second-order perturbation theory~\cite{Rodning:1982zz,Moro:1999fcl,Baur:1977nso,Alder:1956im}, the polarization potential is written as
 \begin{eqnarray}
 V_{\rm pol}=-\frac{1}{2}\alpha\frac{Z^2e^2}{R_d^4(t)}.
 \label{eq:VSOP}
 \end{eqnarray}
 $\alpha$ denotes the {\it E1} polarizability of the deuteron which is defined as~\cite{Alder:1956im}
 \begin{eqnarray}
\alpha &=& \frac{8\pi}{9}\sum_{n\ne0}\frac{B(E1;0\rightarrow n)}{(E_n-E_0)},
\label{eq:alpha}
\end{eqnarray}
 where the indexes $0$ and $n$ denote the ground state and the scattering states of the deuteron, respectively. $B(E1;0\rightarrow n)$ represents the {\it E1} strength for the coupling between the deuteron ground state and the scattering state, which is calculated as follows
 \begin{equation}
B(E1;0\rightarrow n)
=\sum_{M_n,\mu}\left|\left\langle\beta_0,M_0|\mathcal{M}(E1,\mu)|\beta_n,M_n\right\rangle\right|^2.
\label{eq:BE1}
\end{equation}
$|\beta_j\rangle$ denotes the wave function of the deuteron ground state ($j=0$) or the scattering state ($j>0$), which is calculated by solving the following eigenequation:
\begin{eqnarray}
H_{0} |\beta_j \rangle &=& E_{j} \ |\beta_j \rangle \label{eq:EFunc}  ,
\end{eqnarray}
with $E_j$ the eigenenergy of the state $|\beta_j \rangle$.  $H_0$ represents the fully interacting Hamiltonian for the intrinsic motion of the {\it np} system in the absence of any external fields:
\begin{eqnarray}
H_0 &=& T_{\rm rel} + V_{\rm NN}   \label{eq:H0}  ,
\end{eqnarray}
with $T_{\rm rel}$ and $V_{\rm NN}$ being the relative kinetic energy and the {\it NN} interaction, respectively.
$M_0$ and $M_n$ in Eq.~(\ref{eq:BE1}) represent the total angular momentum projections of $|\beta_0\rangle$ and $|\beta_n\rangle$, respectively, while $\mu$ is the angular momentum projection of the {\it E1} operator so that $M_n + \mu = M_0$. In practice we obtain $|\beta_j\rangle$ by diagonalizing the Hamiltonian $H_0$ in the harmonic oscillator (HO) basis, which is characterized by the basis strength $\omega$ and the basis truncation parameter $N_{\rm max}$ (defined as the maximum of twice the radial quantum number plus the orbital angular momentum)~\cite{Barrett:2013nh,Vary:2018jxg}. In the HO basis, the scattering states are discretized. In what follows, we therefore check for the convergence of the scattering observables with increasing basis space cutoff $N_{\rm max}$ for a range of values of $\omega$.
$\mathcal{M}(E1,\mu)$ in Eq.~(\ref{eq:BE1}) is the {\it E1} operator which is expressed in the internal  coordinates of the deuteron as
 \begin{equation}
\mathcal{M}(E1,\mu)
=\sum_{k=n,p}e_kr_kY_{1\mu}(\theta_k,\phi_k).
\label{eq:E1M}
\end{equation}
where $Y_{\lambda\mu}$ denotes the spherical harmonics ($\lambda=1$ denotes the dipole component) in the Condon-Shortley convention. $e_k$ and ${\bm r}_k=(r_k,\theta_k,\phi_k)$ are the charge and the position vector of the proton ($k=p$) or neutron ($k=n$) in the deuteron projectile.

In this work, we adopt five {\it NN} interactions for our investigations: the JISP16~\cite{Shirokov:2005bk,Shirokov:2014kqa}, Daejeon16~\cite{Shirokov:2016ead} {\it NN} interactions and three {\it NN} interactions from the chiral EFT. Specifically, we select the Low Energy Nuclear Physics International Collaboration (LENPIC) {\it NN} interactions~\cite{Epelbaum:2014efa,Epelbaum:2014sza,Binder:2016,Binder:2018} for the chiral EFT interactions, which have been developed for each chiral order up through N$^4$LO. These LENPIC interactions employ a semilocal coordinate-space regulator and we take the interactions with the regulator range of $1.0$ fm. We take three LENPIC interactions at the chiral orders of N$^2$LO, N$^3$LO and N$^4$LO, which we refer to as ``LENPIC-N$^2$LO'', ``LENPIC-N$^3$LO'' and ``LENPIC-N$^4$LO'', respectively.

\subsection{EOM for the scattering}
The full Hamiltonian of the deuteron projectile moving in the time-dependent background field provided by the $^{208}$Pb target can be written as
\begin{eqnarray}
H_{\rm full}(t) &=& H_0 + V_{\rm int}(t) \label{eq:FullH}  ,
\end{eqnarray}
where $V_{\rm int}(t)$ denotes the time-dependent interaction between the projectile and the target during the scattering. The EOM of the projectile during the scattering, in the interaction picture, can be written as
\begin{eqnarray}
i \frac{\partial}{\partial t}|\psi; t \rangle _I &=&  e^{i {H_{0}t}}\ V_{\rm int}(t)\ e^{-i {H_{0}}t}\ |\psi; t \rangle _I \\ \nonumber
&\equiv& V_I(t)\ |\psi; t \rangle _I \ , \label{eq:EOMequation}
\end{eqnarray}
where $V_I(t)$ denotes the time-dependent interaction between the projectile and the target in the interaction picture. The subscript ``I'' specifies the interaction picture. The wave function of the projectile at $t\geq t_0$ can be solved as
\begin{eqnarray}
|\psi; t \rangle_I &=& U_I(t;t_0)|\psi; t_0 \rangle_I , \label{eq:tdwf}
\end{eqnarray}
where $t_0$ denotes the time when the deuteron projectile is in its initial state. $U_I(t;t_0)$ is the unitary operator for the time-evolution
\begin{eqnarray}
U_I(t;t_0) &=& \hat{T}e^{-i\int_{t_0}^t V_I(t')dt'}, \label{eq:uo}
\end{eqnarray}
with $\hat{T}$ the time-ordering operator towards the future.

The numerical solution of the EOM that we employ with the non-perturbative multistep differencing scheme up to the second-order (MSD2) scheme can be found in Refs.~\cite{Yin:2019kqv,Iitaka:1994}.

\subsection{State vector and interaction in the basis representation}
In the numerical implementation of the tBF method, we calculate the time-dependent state vector of the deuteron projectile in a basis representation formed by the deuteron bound and scattering states, i.e., $\{|\beta_j, M_j\rangle\}$. In this basis representation, the state vector of the deuteron projectile at any moment $t$ is written as
\begin{eqnarray}
|\psi; t \rangle_I = \sum_j A_j^I(t)|\beta_j, M_j \rangle \label{eq:tranamp}  ,
\end{eqnarray}
where $A_j^I(t)$ is the amplitude corresponding to the basis $|\beta_j, M_j\rangle$. The time-dependent interaction $V_I(t)$ becomes a matrix in this basis representation. We consider only the {\it E1} component of the Coulomb interaction between the projectile and the target since that is known to be the dominant deuteron excitation mode for sub-barrier scatterings~\cite{Moro:1999fcl}. We calculate the matrix elements of the {\it E1} interaction as follows:
\begin{eqnarray}
&&\langle\beta_j, M_j|V_I(t)|\beta_k, M_k\rangle \\ \nonumber
=& &-\frac{4\pi}{3}Ze^2e^{i(E_j-E_k)t} \sum_\mu\frac{Y_{1\mu}^*(\Omega_R)}{|R_d(t)|^2}\cdot\\ \nonumber
& & \int d \bm{r} \langle\beta_j, M_j|\bm{r}\rangle\frac{r}{2}Y_{1\mu}(\Omega_r)\langle\bm{r}|\beta_k, M_k\rangle\label{eq:E1ele}  ,
\end{eqnarray}
where $\bm{r}=\bm{r}_p-\bm{r}_n$ denotes the relative coordinates of the deuteron projectile.
After adopting an initial state $|\psi; t_0 \rangle_I = \sum_j A_j^I(t_0)|\beta_j, M_j \rangle $ in the above basis representation, we can calculate the state vector of the deuteron projectile at $t>t_0$ with the MSD2 scheme straightforwardly. It is noteworthy that the time-varying $R_d(t)$ (separation between the COMs of the projectile and the target) in Eq.~(\ref{eq:E1ele}) provides the only source of the time-dependence in the present scattering problem.

\subsection{Observables}

%


In this work, we calculate the differential cross-section for elastic scattering by
\begin{equation}
\left(\frac{d\sigma}{d\Omega}\right)_{\rm el} = P_{\rm el}\left(\frac{d\sigma}{d\Omega}\right)_{\rm class}, \label{eq:CSection}
\end{equation}
where the elastic scattering probability $P_{\rm el}$ is obtained by summing over the probabilities in the three orientations of the deuteron ground state after the time-evolution with the tBF method. That is, we take into account the spin flips of the deuteron ground state, a higher-order process, in the elastic channel which would be appropriate for experiments that do not measure the spin of the deuteron in the final state. $\left(\frac{d\sigma}{d\Omega}\right)_{\rm class}$ denotes the classical differential cross-section, which is calculated using a trajectory determined by the adopted potential acting on the COM of the deuteron (either $V_{\rm c}$ or $V_{\rm pot}$). Specifically, we calculate $\left(\frac{d\sigma}{d\Omega}\right)_{\rm class}$ with
\begin{eqnarray}
\left(\frac{d\sigma}{d\Omega}\right)_{\rm class} &=&  \frac{b}{\sin\theta}\left|\frac{db}{d\theta}\right|,
\end{eqnarray}
where $b$ and $\theta$ denote the impact parameter and the scattering angle, respectively. For reference, in the case where $V_{\rm c}$ alone is used, the Rutherford cross-section would emerge since $b=\frac{Ze^2}{2E_d}\cot \left(\frac{\theta}{2}\right)$.

In order to compare with available experimental data~\cite{Rodning:1982zz}, we calculate the following quantity
\begin{eqnarray}
R(E_d)=\frac{\sigma(3\ \rm{MeV}, \theta_1)}{\sigma(3\ \rm{MeV}, \theta_2)}\frac{\sigma(E_d, \theta_2)}{\sigma(E_d, \theta_1)},
\label{eq:RE}
\end{eqnarray}
where $\sigma(E_d,\theta)=2\pi\left(\frac{d\sigma}{d\Omega}\right)_{\rm el}$ denotes the differential cross-section of the elastically scattered deuterons at the angle $\theta$ with the bombarding energy $E_d$.

\section{Results and Discussions}
\label{sec:resultsAndDiscussion}

In this work we set the initial state of the deuteron projectile to be its ground state ($^3S_1-{^3D_1}$ channel). The polarization will be defined for each of the specific applications below. Since {\it E1} transitions respect the conservation of the total spin $S$ of the {\it np} system, we take only channels with $S=1$ into account. We restrict the total angular momentum $J$ to be $J\le 2$ though higher angular momentum states could, in principle, be populated through higher-order transitions. We introduce a quantity $E_{\rm cut}$ to represent the upper energy limit of the retained scattering states of the {\it np} system. To be specific, we adopt the eigenstates of the {\it np} system with eigenenergies below $E_{\rm cut}$ in $^3S_1-{^3D_1}$, $^3P_0$, $^3P_1$, $^3D_2$ and $^3P_2-{^3F_2}$ channels to form the basis representation of the tBF approach in this work.

\begin{figure}[tbh]
\begin{center}
\includegraphics[width=1.0\columnwidth]{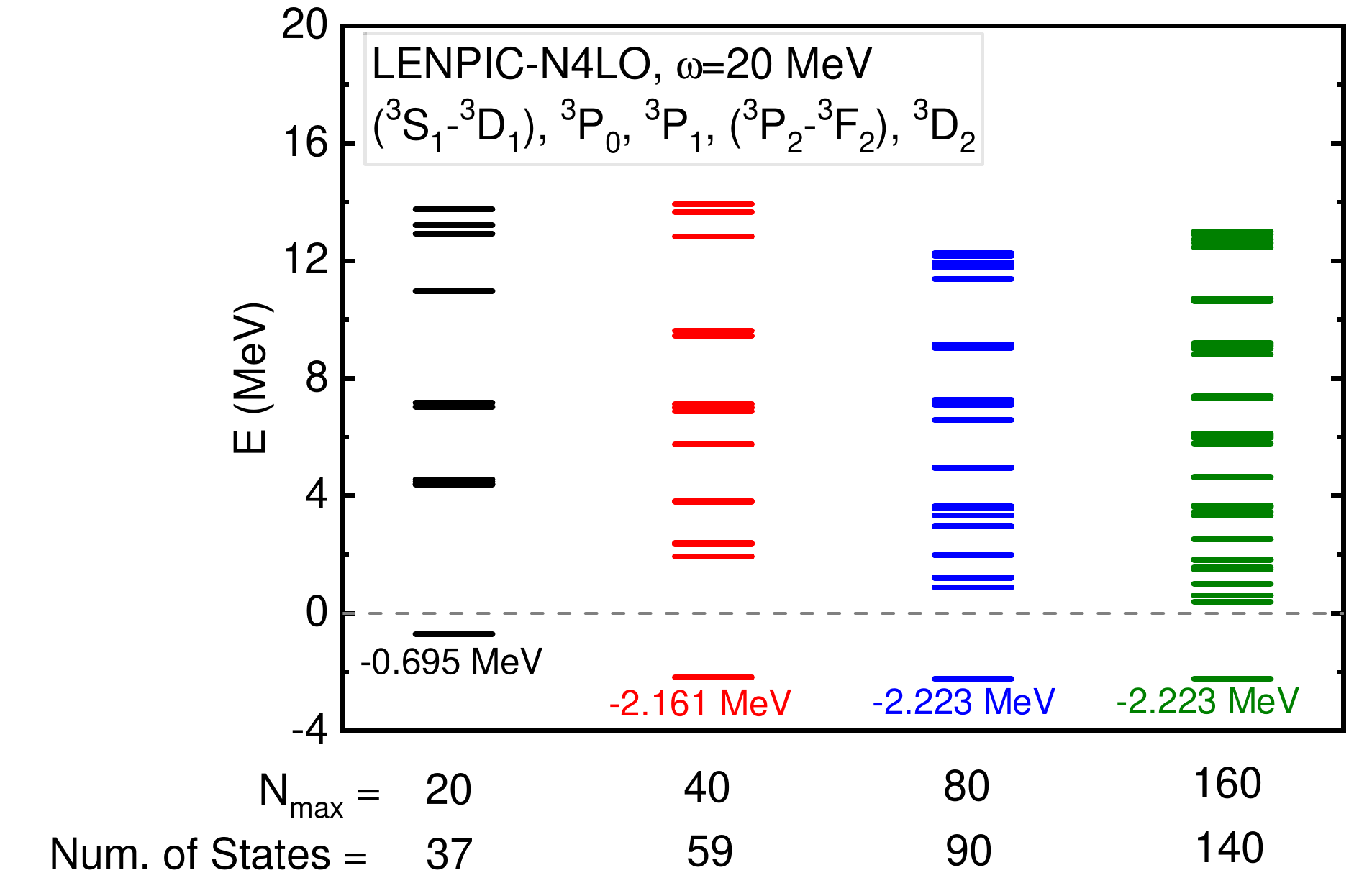}
\end{center}
\caption{(Color online) The deuteron spectrum, up to $E_{\rm cut} = 14$ MeV for $S=1$ and $J\le 2$, as a function of $N_{\rm max}$ obtained with the LENPIC-N$^4$LO interaction with $\omega=20$ MeV. The horizontal dashed line denotes the boundary between the bound and scattering states. We indicate the energy of the deuteron ground state below the lowest level of each spectrum.}\label{fig3}
\end{figure}
In Fig. \ref{fig3}, we show the spectra of the deuteron in five channels ($^3S_1-{^3D_1}$, $^3P_0$, $^3P_1$, $^3P_2-{^3F_2}$, and $^3D_2$) below $14$ MeV calculated with $\omega=20$ MeV and a sequence of $N_{\rm max}$ truncations by adopting the LENPIC-N$^4$LO interaction. We also show the number of states in these five channels which takes into account the degeneracy with respect to the magnetic quantum number of each state. The set of states for a given truncation forms the basis representation of the tBF approach as in Eq.~(\ref{eq:tranamp}) which is used to solve the time-dependent Schr\"{o}dinger equation. The deuteron bound state energy becomes well converged with increasing $N_{\rm max}$ and tends to the experimental value. As the basis truncation $N_{\rm max}$ increases, we find that the discretized scattering states become increasingly dense and flow towards the breakup threshold as expected. The ultimate test of sufficiency of the continuum discretization will be the convergence rate of specific experimental observables with increasing $N_{\rm max}$ as we will address below.

\begin{figure}[tbh]
\begin{flushleft}
\includegraphics[width=1.0\columnwidth]{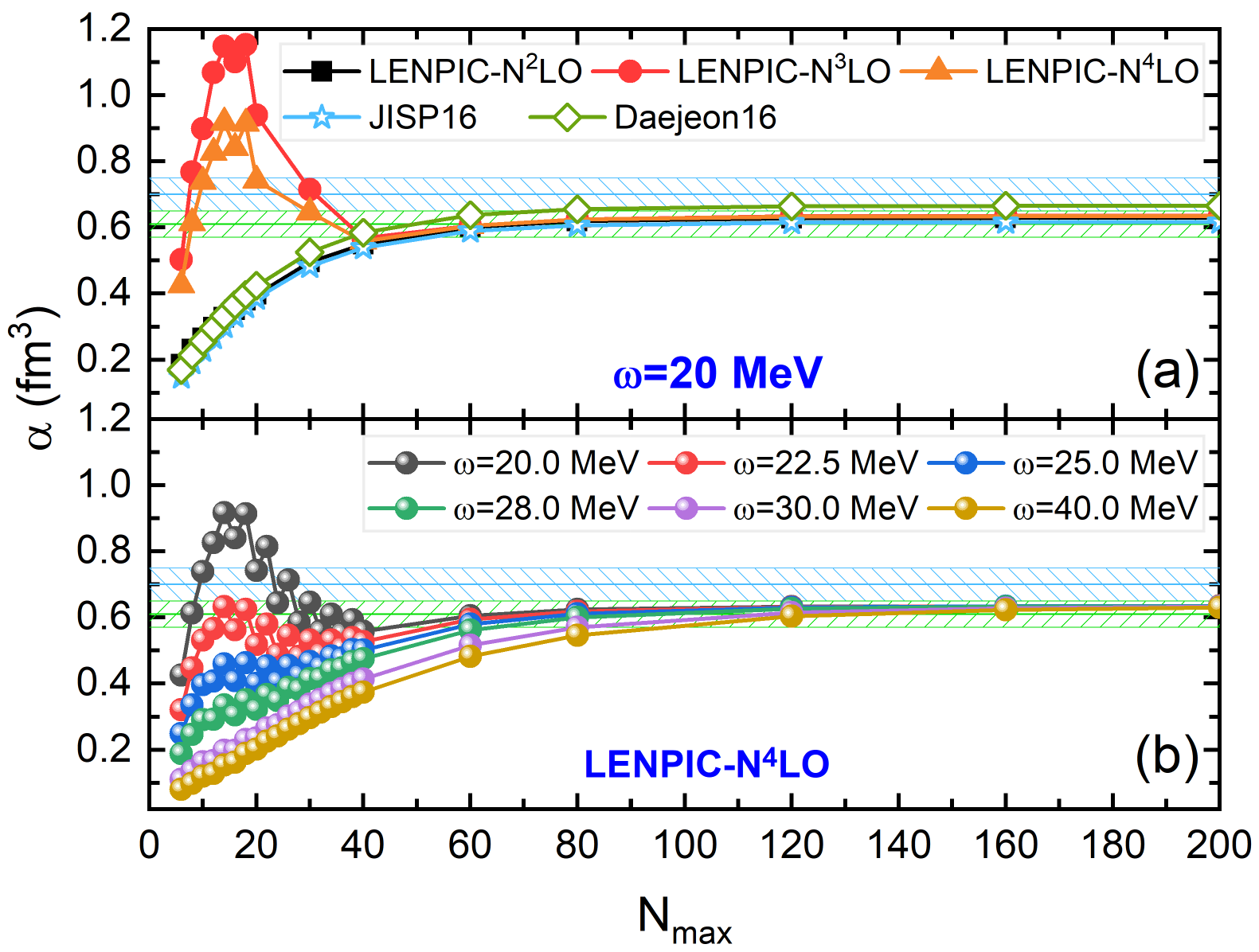}
\end{flushleft}
\caption{(Color online) The {\it E1} polarizability of the deuteron as a function of $N_{\rm max}$. The results in panel (a) are obtained using five different {\it NN} interactions with $\omega=20$ MeV. The results in panel (b) are calculated employing the LENPIC-N$^4$LO interaction with various $\omega$ values. Two experimental results and their uncertainty bands from Ref.~\cite{Rodning:1982zz} (blue region) and Ref.~\cite{Friar:1983zza} (green region) are presented for comparison.}\label{fig4}
\end{figure}
In Fig. \ref{fig4} (a), we show the {\it E1} polarizability of the deuteron, $\alpha$, as a function of the truncation parameter of the HO basis, $N_{\rm max}$, for the LENPIC (-N$^2$LO, -N$^3$LO, -N$^4$LO), JISP16 and Daejeon16 {\it NN} interactions with the identical strength of the HO basis $\omega=20$ MeV. We also show two sets of results extracted from experiments along with their quoted uncertainties for comparison. We find in Fig. \ref{fig4} (a) that the {\it E1} polarizabilities of the deuteron calculated with these five {\it NN} interactions converge to different values at sufficiently large $N_{\rm max}$, which are all consistent with the experimental data. We have extended the calculation to $^4$He polarizability with the {\it ab initio} no-core shell model~\cite{Yin:2024eih}.

The LENPIC-N$^3$LO and LENPIC-N$^4$LO interactions are known to provide challenges for obtaining converged spectra in {\it ab initio} calculations in light nuclei~\cite{Binder:2018}. Since the HO basis space parameters can be characterized by an ultraviolet (UV) cutoff $\Lambda=\sqrt{m_N(N_{\rm max}+3/2)\omega}$ ($m_N$ is the nucleon mass)~\cite{Stetcu:2006ey,Coon:2012ab}, we can deduce from Fig. \ref{fig4} that the {\it E1} polarizability of the deuteron for fixed $\omega$ requires a sufficiently large $N_{\rm max}$ for these interactions to access their higher-momentum contributions. One may interpret the erratic oscillations at small $N_{\rm max}$ for these two interactions in Fig. \ref{fig4} (a) as arising from basis spaces that are deficient in the needed high-momentum contributions. However, the results calculated with the LENPIC-N$^2$LO, JISP16 and Daejeon16 show smooth trends at low to moderate $N_{\rm max}$ indicating that these three interactions are softer in the UV region than the LENPIC-N$^3$LO and LENPIC-N$^4$LO interactions. Above about $N_{\rm max} = 40$ at $\omega = 20$ MeV, we observe that all these five {\it NN} interactions produce a similar trend reflecting similar accumulations of the long-range (infra-red region) components of the nuclear wave function that contribute to the {\it E1} polarizability.

In Fig. \ref{fig4} (b), we present the {\it E1} polarizability of the deuteron as a function of $N_{\rm max}$, calculated with the LENPIC-N$^4$LO interaction with various $\omega$ values. The results with different $\omega$ values converge to the same asymptotic value at sufficiently large $N_{\rm max}$. However, different $\omega$ values result in different convergence patterns at small $N_{\rm max}$. In particular, we observe erratic fluctuations at small $N_{\rm max}$ for small $\omega$ values. However, these fluctuations vanish at large $\omega$ since the UV cutoff $\Lambda$ increases with $\omega$ and the physics at high-momentum is more adequately accommodated at larger $\omega$ and small to moderate $N_{\rm max}$.

\begin{center}
\begin{table}[htb]
\renewcommand\arraystretch{1.3}
\caption{\label{tab:table1} The {\it E1} polarizability of the deuteron, $\alpha$, for $11$ {\it NN} interactions and two results extracted from experiments.}
\setlength{\tabcolsep}{5mm}{
\begin{tabular*}{\linewidth}{l l l}
  \hline\hline
   & $\alpha$ (fm$^3$)  & Refs. \\
   \hline
  LENPIC-N$^2$LO & $0.6292(1)$ & this work \\
  LENPIC-N$^3$LO & $0.6352(1)$ & this work \\
  LENPIC-N$^4$LO & $0.6349(1)$ & this work \\
  JISP16 & $0.6164(1)$ & this work \\
  Daejeon16 & $0.6658(1)$ & this work \\
  Reid soft core (68) & $0.6237$ & \cite{Friar:1997fp} \\
  Reid soft core (93) & $0.6345$ & \cite{Friar:1997fp} \\
  Argonne V$_{14}$ & $0.6419$ & \cite{Friar:1997fp} \\
  Argonne V$_{18}$ & $0.6343$ & \cite{Friar:1997fp} \\
  Bonn (CS) & $0.6336$ & \cite{Friar:1997fp} \\
  Paris & $0.6352$ & \cite{Friar:1997fp} \\
  Exp. & $0.70(5)$ & \cite{Rodning:1982zz} \\
  Exp. & $0.61(4)$ & \cite{Friar:1983zza} \\
  \hline\hline
\end{tabular*}}
\end{table}
\end{center}

We summarize in Table~\ref{tab:table1} the {\it E1} polarizabilities of the deuteron at $N_{\rm max}=240$ obtained with the LENPIC (-N$^2$LO, -N$^3$LO, -N$^4$LO), JISP16 and Daejeon16 interactions with $\omega=20$ MeV. The uncertainty for each interaction denotes the absolute difference between the result at $N_{\rm max}=240$ and the result at $N_{\rm max}=280$. The results calculated with the three LENPIC interactions differ from each other by less than $1\%$. The JISP16 and Daejeon16 interactions predict the smallest and the largest {\it E1} polarizabilities of the deuteron, respectively. The spread of the {\it E1} polarizability, induced by these five interactions, is about $8\%$. For comparison, we also show in Table~\ref{tab:table1} the {\it E1} polarizabilities of the deuteron extracted from experiments in Refs.~\cite{Rodning:1982zz,Friar:1983zza} and those obtained with the Reid soft core (68)~\cite{Reid:1968sq}, Reid soft core (93)~\cite{Stoks:1994wp}, Argonne V$_{14}$~\cite{Wiringa:1984tg}, Argonne V$_{18}$~\cite{Wiringa:1994wb} and Bonn (CS)~\cite{Machleidt:1987hj} interactions in Ref.~\cite{Friar:1997fp}. The results obtained in our work and prior theoretical results~\cite{Friar:1997fp}, all shown in Table~\ref{tab:table1}, are consistent with experiment~\cite{Rodning:1982zz,Friar:1983zza}.

\begin{figure}[htbh]
\centering
\includegraphics[width=1\columnwidth]{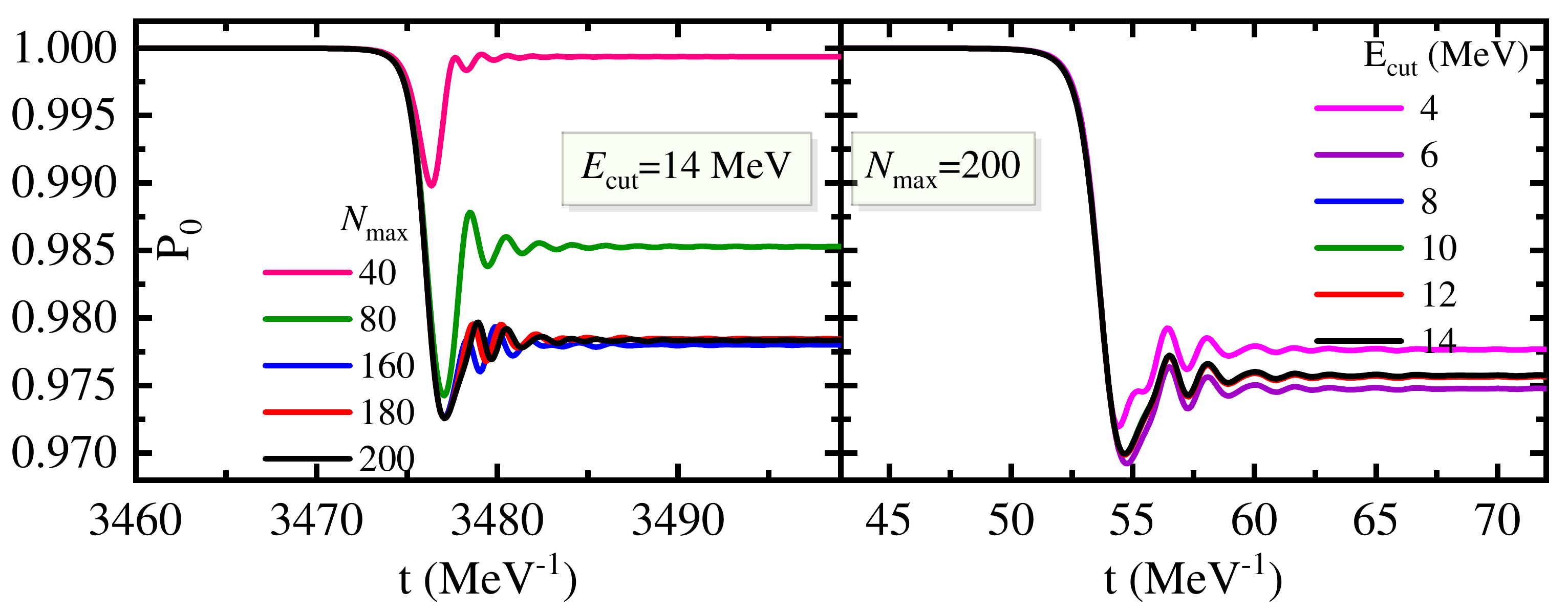}
\caption{(Color online) Probability of the initial state as a function of the evolution time at $E_d=7$ MeV and $\theta=150^\circ$ for various $N_{\rm max}$ [panel (a)] and $E_{\rm cut}$ [panel (b)].}\label{fig5}
\end{figure}

In Fig. \ref{fig5}, we display the probability of the deuteron remaining in its initial state, which we refer to as $P_0$, as a function of the evolution time during the deuteron scattering on $^{208}$Pb at $E_d=7$ MeV and $\theta=150^\circ$, calculated with the tBF method employing the LENPIC-N$^4$LO interaction with $\omega=20$ MeV at various $N_{\rm max}$ values [panel (a)] and various $E_{\rm cut}$ values [panel (b)]. We adopt the trajectory, determined by $V_c+V_{\rm pol}$ [see Eq.~(\ref{eq:Vpot})] with $\alpha=0.635$ fm$^3$. In the following we will use the same $\it {NN}$ interaction and the polarization potential as in Fig. \ref{fig5} unless explicitly stated otherwise. We take a polarized deuteron with the total angular momentum projection $M_0=-1$ in its ground state for the initial state. We show the results for different $N_{\rm max}$ and $E_{\rm cut}$ to investigate the convergence of $P_0$ with respect to these two truncation parameters.

We notice in Fig.~\ref{fig5} that the asymptotic value of the probability $P_0$ is well converged with respect to the truncation parameters $N_{\rm max}$ and $E_{\rm cut}$. The uncertainty of $P_0$ induced by $N_{\rm max}$ and $E_{\rm cut}$ is on the order of $10^{-4}$. In all remaining calculations, we take $N_{\rm max}=200$, $\omega=20$ MeV and $E_{\rm cut}=14$ MeV, which provides $165$ states of the {\it np} system. We deem this to be sufficient since we find the observables that we investigate are converged to the order of $10^{-4}$ under these choices.

\begin{figure}[tbh]
\begin{center}
\includegraphics[width=1.05\columnwidth]{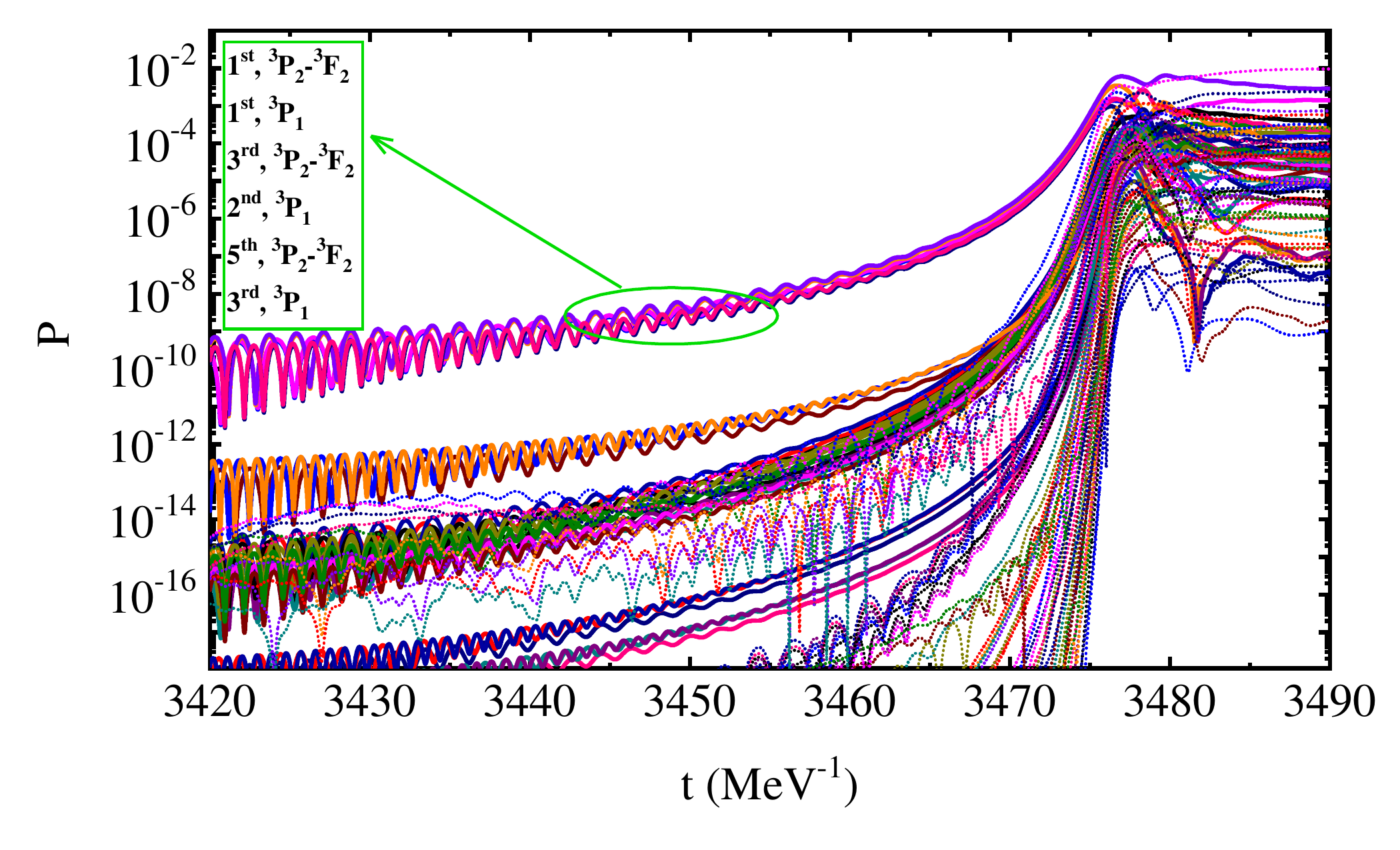}
\end{center}
\caption{(Color online) Probabilities, on a semi-log scale, of the deuteron ground state with $M_0=0$ and $M_0=1$ and the deuteron scattering states as functions of the evolution time for the deuteron scattering on $^{208}$Pb at $E_d=7$ MeV and $\theta=150^\circ$. The solid and the dashed curves represent the states allowed and forbidden respectively by the {\it E1} transition operator acting on the initial state with $M_0=-1$ in first-order perturbation theory. We show in the legend the six states (e.g., ``1st, $^3$P$_2$-$^3$F$_2$'' denotes the lowest state in the $^3$P$_2$-$^3$F$_2$ channel) with the largest probabilities before the crest. These states (all with $M=-1$) are labeled in ascending order of excitation energy in the spectrum of the deuteron.}\label{fig6}
\end{figure}
In Fig. \ref{fig6}, we show the time-evolution of the probabilities of all the states, except for the initial state, for the deuteron scattering on $^{208}$Pb at $E_d=7$ MeV and $\theta=150^\circ$. We take the same initial state (i.e., a polarized deuteron ground state with $M_0=-1$) as in Fig. \ref{fig5}. We signify the states, allowed and forbidden by the {\it E1} transition operator acting on our initial state in first-order perturbation theory, by the solid and the dashed curves in Fig. \ref{fig6}, respectively. For simplicity, we will refer to these states as ``{\it E1} allowed'' and ``{\it E1} forbidden'', respectively.

In the early stage of the time-evolution in Fig. \ref{fig6}, only {\it E1} allowed states populate significantly. We note that six of them (see the legend of Fig. \ref{fig6}) populate well above the others before reaching the crest. The wave functions of the first, third and fifth states in the $^3P_2-^3F_2$ channel are dominated by the $^3P_2$ component and therefore obtain significant populations from the $^3S_1$ component of the deuteron ground state in this stage, according to the {\it E1} selection rules. However, the second and fourth states in the $^3P_2-^3F_2$ channel, which are dominated by the $^3F_2$ channel, are not able to populate via direct transitions from the $^3S_1$ component of the deuteron ground state and therefore are less populated than the first, third and fifth states in this channel. The populations of the {\it E1} forbidden states are very small, but still non-zero, during this early stage. Of course, {\it E1} forbidden states populate through transitions from the {\it E1} allowed states according to the {\it E1} selection rule. Therefore, as expected, the {\it E1} forbidden states do not populate significantly until the {\it E1} allowed states accumulate appreciable populations. Note that the de-excitations among all the states are also taken into account in the tBF method. One may be able to visualize that an entire complex web of quantum excitations and de-excitations is evolving with time and Fig. \ref{fig6} attempts to portray some aspects of the probability flow among these states.

\begin{figure}[tbh]
\begin{center}
\includegraphics[width=1.0\columnwidth]{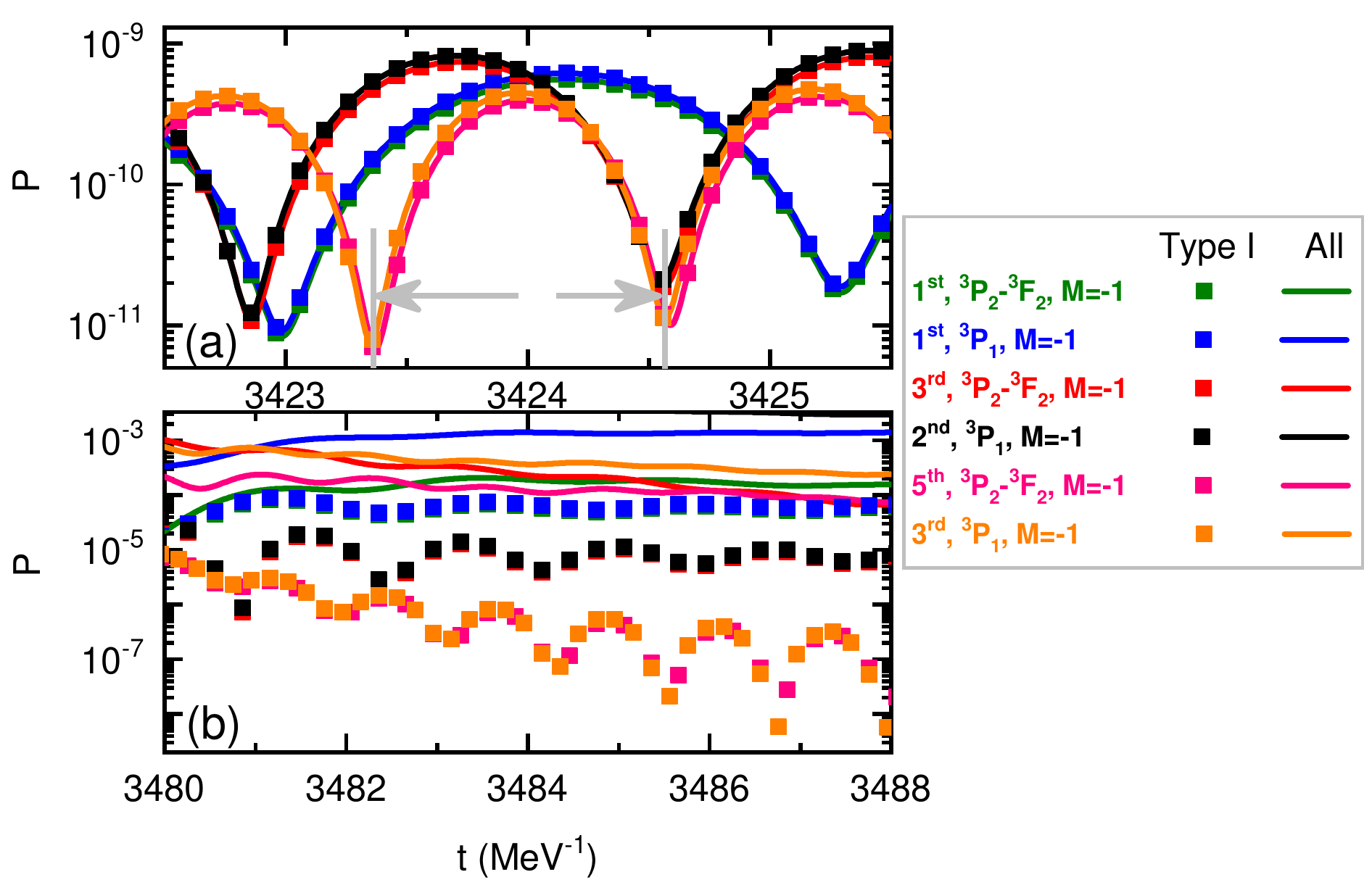}
\end{center}
\caption{(Color online) Time-evolution for the probabilities of six selected states in the incoming [panel (a)] and outgoing [panel (b)] segments of the d+$^{208}$Pb scattering at $E_d=7$ MeV and $\theta=150^\circ$, which are calculated with all the couplings (solid curves) or Type \uppercase\expandafter{\romannumeral1} coupling (solid squares). These states are labeled in ascending order of excitation energy in the spectrum of the deuteron. The horizontal arrows delimit a cycle that is discussed in the text.}\label{fig9}
\end{figure}
We note that the probabilities of all the states in Fig. \ref{fig6} exhibit oscillations. To shed more light on these oscillating patterns, we show in Fig. \ref{fig9} the probabilities of the six states, highlighted in Fig. \ref{fig6}, as an example. In panel (a) we show the results in the early stage of the time-evolution, which we refer to as the ``incoming segment'', in an expanded view. In panel (b) we show the results after the closest approach, which we refer to as the ``outgoing segment'', also in an expanded view. For both the incoming and outgoing segments, we present in Fig. \ref{fig9} the results calculated by the tBF method with all the couplings and with Type \uppercase\expandafter{\romannumeral1} coupling which represents the case where we retain only the couplings between the initial state and its {\it E1} allowed states. In the incoming segment, we notice that the results for both of the two sets of couplings are coincident since these are six states that populate dominantly in the incoming segment as shown in Fig. \ref{fig6}. As the deuteron projectile approaches the $^{208}$Pb target, the higher-order effects become significant. As a consequence, the results from these two couplings become significantly different in the outgoing segment.
\begin{center}
\begin{table*}[!htpb]
\renewcommand\arraystretch{1.0}
\caption{\label{tab:table2} $E_x$, $\Gamma$ and $\kappa$ for the six scattering states of the deuteron which are shown in Fig. \ref{fig9}. $\Gamma$ and $\kappa$ in the incoming segment are distinguished from those in the outgoing segment for the tBF calculations with either the Type \uppercase\expandafter{\romannumeral1} coupling or all the couplings. See the text for additional details.}
\setlength{\tabcolsep}{4mm}{
\begin{tabular}{|l|c|c|c|c|c|c|c|c|c|}
\hline
\multirow{3}{*}{ } & \multirow{3}{*}{$E_{\rm x}$ (MeV)} & \multicolumn{4}{c|}{$\Gamma$(MeV$^{-1}$)} & \multicolumn{4}{c|}{$\kappa$}\\
\cline{3-10}
 & & \multicolumn{2}{c|}{Incoming} & \multicolumn{2}{c|}{Outgoing} & \multicolumn{2}{c|}{Incoming}& \multicolumn{2}{c|}{Outgoing}\\
 \cline{3-10}
  & & Type \uppercase\expandafter{\romannumeral1} & All  & Type \uppercase\expandafter{\romannumeral1} & All & Type \uppercase\expandafter{\romannumeral1} & All & Type \uppercase\expandafter{\romannumeral1} & All \\
\hline
$1^{\rm st}$ $^3$P$_2$-$^3$F$_2$& $2.72$ & $2.31$ & $2.31$ & $2.30$ & $2.10$ & $6.28$ & $6.28$ & $6.26$ & $5.71$\\
$1^{\rm st}$ $^3$P$_1$& $2.72$ & $2.31$ & $2.31$ & $2.30$ & $2.13$ & $6.28$ & $6.28$ & $6.26$ & $5.79$\\
$3^{\rm rd}$ $^3$P$_2$-$^3$F$_2$& $3.69$ & $1.70$ & $1.70$ & $1.69$ & $2.16$ & $6.27$ & $6.27$ & $6.24$ & $7.97$\\
$2^{\rm nd}$ $^3$P$_1$& $3.70$ & $1.70$ & $1.70$ & $1.69$ & $2.12$ & $6.29$ & $6.29$ & $6.25$ & $7.84$\\
$5^{\rm th}$ $^3$P$_2$-$^3$F$_2$& $5.15$ & $1.22$ & $1.22$ & $1.17$ & $1.32$ & $6.28$ & $6.28$ & $6.03$ & $6.80$\\
$3^{\rm rd}$ $^3$P$_1$& $5.17$ & $1.22$ & $1.22$ & $1.17$ & $1.31$ & $6.31$ & $6.31$ & $6.05$ & $6.77$\\
\hline
\end{tabular}}
\end{table*}
\end{center}
We find in Fig. \ref{fig9} that the probabilities of all these six states oscillate in approximately periodic patterns in each segment for both of the two sets of couplings. In Fig. \ref{fig9} we define a quantity $\Gamma$, for every state in a given segment, as the time difference between two adjacent troughs, within the time range of the same segment. We show in Table~\ref{tab:table2} the quantity $\Gamma$ of these six states in both the incoming  (Columns 3 and 4) and the outgoing segments (Columns 5 and 6). We also distinguish in Table~\ref{tab:table2} the quantity $\Gamma$ calculated with the Type \uppercase\expandafter{\romannumeral1} coupling (Columns 3 and 5) from the one with all the couplings (Columns 4 and 6). We then calculate the products of the excitation energy $E_x$ (Column 2) and $\Gamma$, denoted as $\kappa$, and present the results in Table~\ref{tab:table2} (Columns 7-10). We find in Table~\ref{tab:table2} that the $\kappa$ values calculated with these two couplings (Columns 7 and 8) are the same up to the third significant figure for each state in the incoming segment. This is readily understood since the higher-order effects are negligible in this segment. We also notice in Table~\ref{tab:table2} that the $\kappa$ values of these six states are all approximately $2\pi$. We recognize this as a ``phase coherence'' representing periodicity in the entanglement of the ground state with these particular excited states induced by the {\it E1} operator.

In the outgoing segment, we find that the $\kappa$ values of these six states, calculated with the Type \uppercase\expandafter{\romannumeral1} coupling (Column 9), are also around $2\pi$, which means the phase coherence phenomenon remains valid for this case. However, we observe noticeable differences for the $\kappa$ values of these six states, calculated with all the couplings (Column 10), and most of the $\kappa$ values deviate significantly from $2\pi$. We refer to this as a ``phase decoherence''. Thus, the maintenance of the phase coherence phenomenon depends on the absence of the couplings between the excited states. Such interference effect at the amplitude level due to the time-varying interaction between the particle and the external field is reminiscent of the Landau-Pomeranchuk-Migdal (LPM) effect known in relativistic inelastic collisions~\cite{Landau:1953um,Migdal:1956tc,Yao:2022eqm}.

\begin{figure}[htbh]
\begin{center}
\includegraphics[width=1.0\columnwidth]{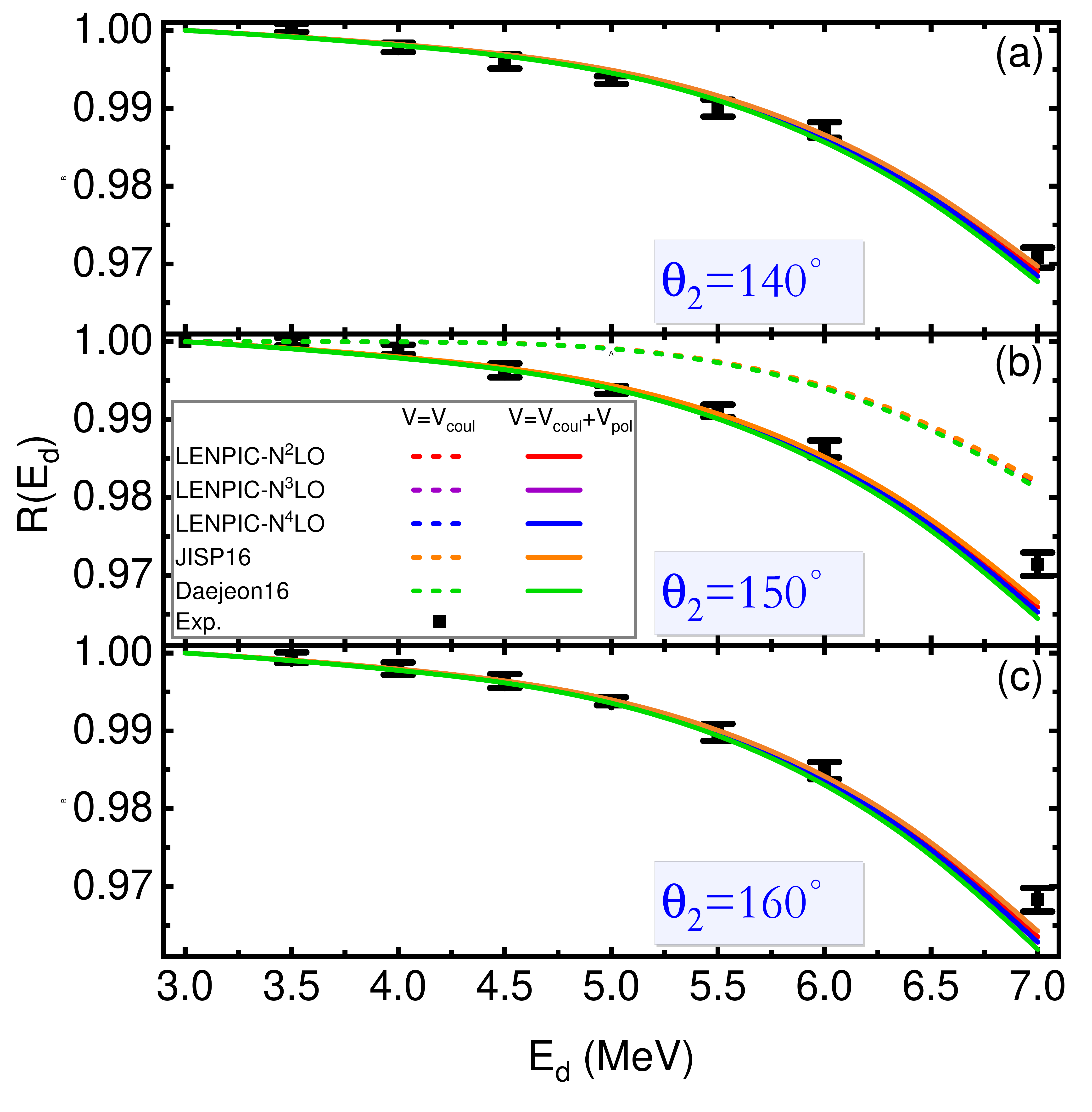}
\end{center}
\caption{(Color online) $R(E_d)$ with $\theta_2=140^\circ$ [panel (a)], $150^\circ$ [panel (b)] and $160^\circ$ [panel (c)]as a function of the bombarding energy $E_d$ for the scattering of d+$^{208}$Pb. The solid lines represent the tBF results employing five {\it NN} interactions including their associated polarization potentials. The dashed lines in panel (b) denote the results calculated with no polarization potential.  The experimental data (black solid squares with error bars) are also shown for comparison.}\label{fig11}
\end{figure}
In Fig. \ref{fig11} we show the quantity $R(E_d)$ with $(\theta_1,\theta_2)=(60^\circ,140^\circ)$, $(60^\circ,150^\circ)$ and $(60^\circ,160^\circ)$ for the scattering of the deuteron on $^{208}$Pb at $E_d=3-7$ MeV. We omit $\theta_1$ in the following since we use the same $\theta_1=60^\circ$. We adopt the initial state with equal amplitudes for the three magnetic substates. We also present the experimental data in Fig. \ref{fig11} for comparison~\cite{Rodning:1982zz}. To investigate the sensitivity of $R(E_d)$ to the {\it NN} interactions, we show in Fig. \ref{fig11} the tBF results calculated with the LENPIC (-N$^2$LO, -N$^3$LO, -N$^4$LO), JISP16 and Daejeon16 {\it NN} interactions. For pure Rutherford scattering $\sigma(E_d,\theta_1)/\sigma(E_d,\theta_2)$ in Eq.~(\ref{eq:RE}) is independent of $E_d$ and therefore $R(E_d)=1$ in this case~\cite{Rodning:1982zz,Moro:1999fcl}.  Therefore the deviation of the quantity $R(E_d)$ from unity, i.e., $1-R(E_d)$, reveals the deviation of a scattering from the classical Rutherford scattering.

In order to study the effect of the polarizabition potential separately, we present in Fig. \ref{fig11} (b) the tBF results without the polarization potential (dashed lines) for $\theta_2=150^\circ$ for example. That is, we use the Rutherford trajectories determined by the Coulomb potential. $1-R(E_d)$ is entirely induced by the internal transitions in the deuteron projectile in this case.
We notice in Fig. \ref{fig11} (b) that the calculations with no polarization potential are not able to reproduce the experimental data. By taking the polarization potential into account (solid lines), our tBF approach reproduces the experimental data at $E_d=3-6$ MeV while falling below experiment at $E_d=7$ MeV for all the five adopted {\it NN} interactions, which indicates that the polarization potential is significant in the current application. In our calculations we take $\alpha$ from Table~\ref{tab:table1} and retain three significant figures. In Ref.~\cite{Yin:2019kqv}, we have compared the tBF results of the LENPIC-N$^4$LO interaction with the optical model calculations in Refs.~\cite{Rodning:1982zz,Moro:1999fcl}. We will concentrate mainly on the sensitivity of $R(E_d)$ to the {\it NN} interactions in the following.

We find in Fig. \ref{fig11} that all the adopted five interactions are able to reproduce the experimental $R(E_d)$ at $E_d=3-6$ MeV. $R(E_d)$ falls below the experimental data at $E_d=7$ MeV for $\theta_2=150^\circ$ and $160^\circ$ which we attribute to neglecting the effect of the energy loss (due to the internal excitations) of the projectile on its center of mass motion in the tBF method. $R(E_d)$ with $\theta_2=140^\circ$ at $E_d=7$ MeV is close to experiment since the energy loss is less significant compared to $\theta_2=150^\circ$ and $160^\circ$. The energy transferred to the intrinsic degree of freedom of the projectile, approximated by the average excitation energy of the projectile, should be compensated by a reduction of the energy in its COM degree of freedom. We have shown in Ref.~\cite{Yin:2019kqv} that the energy loss effect could enhance $R(E_d=7$ MeV$)$ by about $1\%$ using the LENPIC-N$^4$LO interaction, which would lead to an improved tBF description of the experimental data at $E_d=7$ MeV.
The effect of the energy loss on $R(E_d)$ is insignificant at $E_d=3-6$ MeV and therefore the current calculation well reproduces those experimental data.

%

\begin{figure}[tbh]
\begin{center}
\includegraphics[width=1.05\columnwidth]{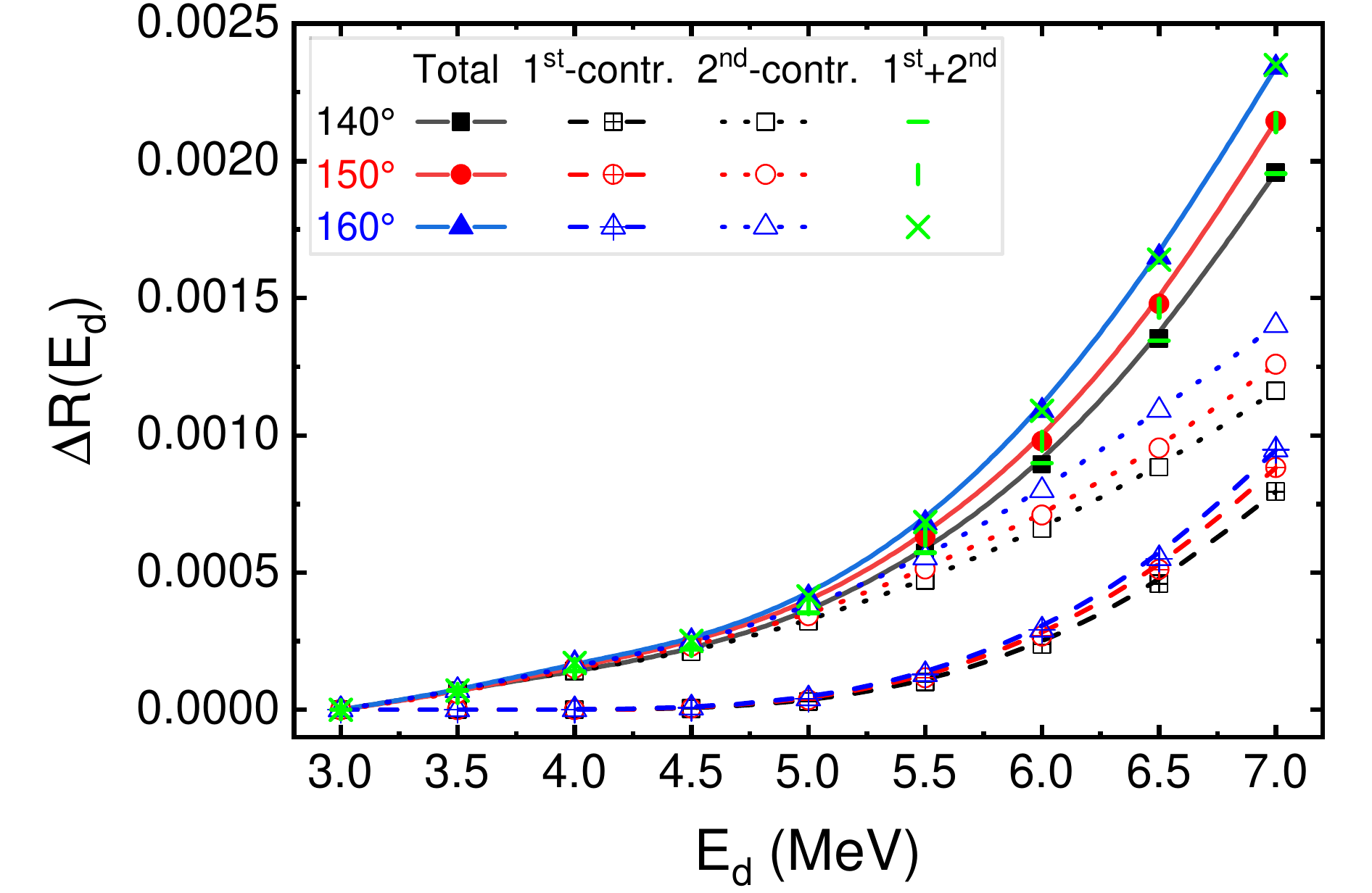}
\end{center}
\caption{(Color online) Contributions to the spread of $R(E_d)$ arising from different {\it NN} interactions.  The total spread (solid symbols and solid lines) is depicted for three scattering angles ($\theta_2=140^\circ, 150^\circ$ and $160^\circ$). In particular, we show the contributions to the total spread at each angle arising from the {\it E1} transition matrix elements (dashed line+symbol embedded with `+') and from the {\it E1} polarizability of the deuteron (dotted line+open symbol).
The black, red and blue colors represent the results for $\theta_2=140^\circ, 150^\circ$ and $160^\circ$, respectively.
The green symbols `$-$ $|$ $\times$' denote the addition of these two contributions for $\theta_2=140^\circ, 150^\circ$ and $160^\circ$, respectively. The deviation of the green symbols from their solid symbol counterparts is small and hardly visible on this scale (see the text for details).}\label{fig13}
\end{figure}

We find in Fig. \ref{fig11} that $R(E_d)$ calculated with the five interactions are almost indistinguishable at low incident energies and separate with increasing incident energy. For deeper insights into the sensitivity of $R(E_d)$ to {\it NN} interactions, we present in Fig. \ref{fig13} the spread of $R(E_d)$, $\Delta R(E_d)$, evaluated by the absolute difference between the upper and lower edges of the solid lines in Fig. \ref{fig11}, as a function of the deuteron incident energy (denoted as ``Total'' in Fig. \ref{fig13}). We notice in Fig. \ref{fig13} that the spread $\Delta R(E_d)$ increases with the deuteron incident energy as well as the scattering angle, indicating that $R(E_d)$ is more sensitive to {\it NN} interaction at higher incident energies and larger scattering angles.

In this work, $\Delta R(E_d)$ stems from the following two effects. The first effect is the dependence of the {\it E1} transition matrix elements [Eq.~(\ref{eq:E1ele})] on {\it NN} interactions, which contribute to $\Delta R(E_d)$ via the time-dependent Schr\"{o}dinger equation [Eq.~(\ref{eq:EOMequation})]. The second effect is the dependence of the {\it E1} polarizability of the deuteron on the {\it NN} interactions, which contributes to $\Delta R(E_d)$ via the effect of the polarization potential on the scattering trajectory. Note that $\Delta R(E_d)$ is not simply a linear superposition of the above two effects. We therefore perform the following two sets of independent tBF calculations to investigate their individual contributions to the total $\Delta R(E_d)$ in Fig. \ref{fig13}. To evaluate the contribution of the first effect, we solve the time-dependent Schr\"{o}dinger equation using different {\it E1} transition matrix elements calculated with the five interactions, where the trajectories are determined by the Coulomb potential and the polarization potential with the same {\it E1} polarizability $\alpha=0.635$ fm$^3$, for each $\theta_2$. We present in Fig. \ref{fig13} the corresponding spread of $R(E_d)$, which is the absolute difference between the two extreme values of $R(E_d)$ obtained with these five interactions at each $E_d$ and $\theta_2$, denoted as ``1$^{\rm st}$-contr.''. To estimate the contribution of the second effect, we solve the time-dependent Schr\"{o}dinger equation using the same {\it E1} transition matrix elements of the LENPIC-N$^4$LO interaction, while we use the trajectories, determined by the Coulomb potential and the polarization potential with the two extreme values of the five {\it E1} polarizabilities obtained in this work (Table~\ref{tab:table1}), i.e., $\alpha=0.616$ fm$^3$ and $\alpha=0.666$ fm$^3$, for each $\theta_2$. We show in Fig. \ref{fig13} the corresponding spread of $R(E_d)$, which stems entirely from using these two extreme {\it E1} polarizabilities, denoted as ``2$^{\rm nd}$-contr.''. We find in Fig. \ref{fig13} that the addition of these two contributions, denoted as ``1$^{\rm st}$+2$^{\rm nd}$'', can almost reproduce the total $\Delta R(E_d)$, which indicates that the above evaluations to the individual contributions of these two effects appear to be reasonable.

At low incident energies, since the internal transitions in the deuteron projectile are insignificant and the polarization potential is weak, we find in Fig. \ref{fig13} that both the total spread of $R(E_d)$ and its two contributions are negligibly small. Both of these two effects grow as the incident energy increases. Therefore the spreads of $R(E_d)$ induced by these two effects and the total spread become more noticeable at higher incident energies. We find the spread of $R(E_d)$ stems almost completely from the second effect at low incident energies. As the incident energy increases, the second effect remains dominant, though the first effect is growing in the direction of becoming comparable.

One should note that all the interactions adopted in this work describe the {\it NN} scattering phase shifts accurately and therefore are approximately equivalent on-shell. This suggests that the increasing sensitivity of $R(E_d)$ with increasing deuteron incident energy stems from the differing off-shell properties of these interactions. Therefore, investigating d+$^{208}$Pb scattering at higher energies as well as other nuclear reactions with the tBF approach is a promising approach to constrain the off-shell behavior of {\it NN} interactions in the future.

\begin{center}
\begin{table}[!htpb]
\renewcommand\arraystretch{1.5}
\caption{\label{tab:table3} $\chi^2/{\rm data}(\theta_2)$ and $\chi^2_t/{\rm data}$ for the five {\it NN} interactions presented in Fig. \ref{fig11}. See the text for details.}
\setlength{\tabcolsep}{1.2mm}{
\begin{tabular*}{\linewidth}{l c c c c}
  \hline\hline
   & \thead[l]{$\chi^2$/data \\$\theta_2=140^\circ$}  & \thead[l]{$\chi^2$/data \\$\theta_2=150^\circ$} & \thead[l]{$\chi^2$/data \\$\theta_2=160^\circ$}  & $\chi_t^2$/data \\
   \hline
  LENPIC-N$^2$LO & $2.703$ & $3.105$ & $1.602$ & $2.470$\\
  LENPIC-N$^3$LO & $2.778$ & $3.619$ & $2.053$ & $2.817$\\
  LENPIC-N$^4$LO & $2.789$ & $3.593$ & $2.036$ & $2.806$\\
  JISP16         & $2.725$ & $2.657$ & $1.210$ & $2.197$\\
  Daejeon16      & $2.837$ & $4.517$ & $2.829$ & $3.395$\\
  \hline\hline
\end{tabular*}}
\end{table}
\end{center}

To quantify the deviation of our tBF results of $R(E_d)$ from the experimental data for $\theta_2=140^\circ, 150^\circ$ and $160^\circ$, we present in Table~\ref{tab:table3} (Column $2-4$) the quantity $\chi^2/{\rm data}(\theta_2)$, which we define as follows
\begin{eqnarray}
\chi^2/{\rm data}(\theta_2)=\frac{1}{7}\sum_{i=1}^{7}\left[\frac{R^{\rm tBF}_i(E_d)-R^{\rm Exp}_i(E_d)}{\delta R^{\rm Exp}_i(E_d)}\right]^2,
\label{eq:chisquare}
\end{eqnarray}
where $R^{\rm Exp}_i(E_d)$ denotes the central value of the experimental $R(E_d)$ in Fig. \ref{fig11}. $R^{\rm tBF}_i(E_d)$ represents the tBF results for each interaction. $\delta R^{\rm Exp}_i(E_d)$ corresponds to the experimental error. The subscript $i$ runs over the seven points experimentally available for each $\theta_2$. To evaluate the entire deviation of our results from the experimental data, we also present in Table~\ref{tab:table3} the quantity $\chi_t^2/{\rm data}$ (Column $5$), defined by
\begin{eqnarray}
\chi_t^2/{\rm data}=\frac{1}{21}\sum_{j=1}^{21}\left[\frac{R^{\rm tBF}_j(E_d)-R^{\rm Exp}_j(E_d)}{\delta R^{\rm Exp}_j(E_d)}\right]^2,
\label{eq:chisquare}
\end{eqnarray}
where the subscript $j$ runs over all the $21$ points experimentally available in Fig. \ref{fig11}.

We find in Table~\ref{tab:table3} that $\chi^2/{\rm data}(\theta_2)$ becomes more sensitive to {\it NN} interaction as $\theta_2$ increases, which is consistent with the results in Fig. \ref{fig13}. All the five interactions adopted in this work provide rather small $\chi_t^2/{\rm data}$, indicating that the tBF approach is able to describe reasonably the deuteron scattering on $^{208}$Pb at $E_d=3-7$ MeV with these interactions.
We expect to improve $\chi_t^2/{\rm data}$ by taking into account the correction of the energy loss to the COM of the projectile in the future.

\section{Conclusions and outlook}
\label{sec:conclusions}
In this work, we investigated the scattering of the deuteron projectile on the $^{208}$Pb target below the Coulomb barrier with the non-perturbative time-dependent basis function (tBF) approach, employing the LENPIC (-N$^2$LO, -N$^3$LO, -N$^4$LO), JISP16 and Daejeon16 nucleon-nucleon ({\it NN}) interactions. We constructed the basis representation of the tBF method with the deuteron ground and discretized scattering states, which are obtained by diagonalizing the realistic Hamiltonian based on these interactions in a large harmonic oscillator (HO) basis sufficient to obtain the observables presented here. We considered all the possible electric-dipole ({\it E1}) transitions among these states during the scattering. We employed the classical trajectory determined by either the Coulomb potential or the Coulomb potential supplemented with the polarization potential. For each {\it NN} interaction that we employed, we adopted the {\it E1} polarizability of the deuteron $\alpha$ (in the polarization potential) obtained with the same interaction.
For each of our chosen {\it NN} interactions, the tBF calculations have no adjustable parameters.

Before performing the tBF calculations, we studied the spectra and {\it E1} polarizability of the deuteron projectile, which are both time-independent. Taking the LENPIC-N$^4$LO interaction as an example, we found that the deuteron ground state energy becomes well converged and tends to the experimental value with increasing $N_{\rm max}$. By comparing the {\it E1} polarizabilities of the deuteron, calculated with the five adopted interactions, we found that all these five interactions produce converged results at sufficiently large $N_{\rm max}$, which are all consistent with the data extracted from experiments. The theoretical {\it E1} polarizabilities show significant interaction dependence with the spread of about $8\%$.

We then investigated the scattering dynamics during the scattering of d+$^{208}$Pb at $E_d=7$ MeV and $\theta=150^\circ$ with the LENPIC-N$^4$LO interaction. The asymptotic value of the probability for the deuteron remaining in its initial state is well converged for our chosen values of the two truncation parameters $N_{\rm max}$ and $E_{\rm cut}$. The {\it E1} allowed states, especially six of the scattering states, populate dominantly in the early stage of the scattering and the {\it E1} forbidden states populate significantly in the later scattering process via the higher-order effects.

By investigating the oscillations of the probabilities of these six scattering states mentioned above with the Type \uppercase\expandafter{\romannumeral1} coupling (only allowed transitions from the ground state retained) as well as with all the possible couplings, we established a phase coherence in the incoming segment representing {\it E1}-induced entanglement of each of these six states with the initial state along with probability oscillations commensurate with their energy differences. However, we observed phase decoherence for the results with all the possible couplings included, due to the higher-order effects in the outgoing segment reminiscent of the Landau-Pomeranchuk-Migdal (LPM) effect. This was confirmed by showing that phase coherence is retained in the outgoing segment with the Type \uppercase\expandafter{\romannumeral1} coupling.

Finally, we investigated the elastic cross-section ratios $R(E_d)$. By considering the internal {\it E1} transitions in the deuteron projectile and the polarization potential during the scattering, the tBF approach provides overall good descriptions to the available experimental data at $E_d\le7$ MeV. $R(E_d)$ becomes more sensitive to {\it NN} interactions with increasing either the deuteron incident energy or the scattering angle $\theta_2$. In particular, the sensitivity of $R(E_d)$ to {\it NN} interactions stems almost completely from the dependence of the {\it E1} polarizability on the {\it NN} interactions at low incident energies. Its contribution to the spread remains dominant in contrast to the contribution of the dependence of the {\it E1} transition matrix elements on the {\it NN} interactions, though both of these two contributions grow with increasing incident energy.

In the future, we will improve the tBF method to extend the range of its applications. (1) We will extend the tBF method to a fully quantum mechanical framework, within which the center of mass motion and the internal motion of the projectile will be considered coherently. The energy loss omitted in this work will then be resolved naturally. (2) We are extending the tBF method to investigate nuclear reactions involving larger projectile nuclei, where we will obtain the nuclear structure information with {\it ab initio} approaches, e.g., no-core shell model. (3) In the tBF method we will introduce the chiral two-body charge operator, which first appears at N$^3$LO, as a higher-order correction to the present one-body {\it E1} operator~\cite{Filin:2019eoe}. We will also investigate the contribution of other electromagnetic transitions, such as {\it E2} and {\it M1}, to scattering observables. (4) We will extend the tBF method to higher bombarding energies by considering the strong interaction between the projectile and the target~\cite{Satchler:1979ni,Moffa:1977zz}.

\section*{Acknowledgments}
We acknowledge helpful discussions with Andrey Shirokov, Pieter Maris, Shiplu Sarker, Robert A. M. Basili, Antonio M. Moro, Gerhard Baur, Zhigang Xiao, Li Ou, William Lynch and Betty Tsang.
This work is supported in part by the U.S. Department of Energy (DOE) under Grant DE-SC0023692.
A portion of the computational resources are provided by the National Energy Research Scientific Computing Center (NERSC), a U.S. Department of Energy Office of Science User Facility located at Lawrence Berkeley National Laboratory, operated under Contract No. DE-AC02-05CH11231 using NERSC award NP-ERCAP0020944. Xingbo Zhao is supported by new faculty startup funding by the Institute of Modern Physics, Chinese Academy of Sciences, by Key Research Program of Frontier Sciences, Chinese Academy of Sciences, Grant No. ZDB-SLY-7020, by the Natural Science Foundation of Gansu Province, China, Grant No. 20JR10RA067, by Gansu International Collaboration and Talents Recruitment Base of Particle Physics (2023-2027), by International Partnership Program of the Chinese Academy of Sciences, Grant No. 016GJHZ2022103FN, by National Natural Science Foundation of China, Grant No. 12375143, and by National Key R\&D Program of China, Grant No. 2023YFA1606903.
Peng Yin and Wei Zuo are supported by the National Natural Science Foundation of China (Grant Nos. 11975282, 11705240, 11435014), the Strategic Priority Research Program of Chinese Academy of Sciences, Grant No. XDB34000000, the Natural Science Foundation of Gansu Province under Grant No. 23JRRA675, and the Key Research Program of the Chinese Academy of Sciences under Grant No. XDPB15.
This work is also partially supported by Shanghai Research Center for
Theoretical Nuclear Physics, NSFC and Fudan University, Shanghai 200438, China, the National Natural Science Foundation of China under Grant No.12147101, and the CUSTIPEN (China-U.S. Theory Institute for Physics with Exotic Nuclei) funded by the U.S. Department of Energy, office of Science under Grant No. DE-SC0009971.

\end{document}